%
%
%
%
\magnification 1175
\baselineskip 13 pt
 \def\ol{\overline}
 \def\1i{\'{\i}}
 
 \def\ie{{i.e.}}
 \def\Par{\par\vskip 5 pt}
 \def\eqn#1{ \eqno({#1}) \qquad }
 \def\2n{\~n}
 \def\ni{\noindent}
\newbox\Ancha
\def\gros#1{{\setbox\Ancha=\hbox{$#1$}
   \kern-.025em\copy\Ancha\kern-\wd\Ancha
   \kern.05em\copy\Ancha\kern-\wd\Ancha
   \kern-.025em\raise.0433em\box\Ancha}}
%
 %
 \def\er{{\bf\hat e}_{r}}
 \def\eth{{\bf\hat e}_{\theta}}
 \def\ez{{\bf\hat e}_{z}}


 %
\font\bigggfnt=cmr10 scaled \magstep 3
\font\biggfnt=cmr10 scaled \magstep 2
\font\bigfnt=cmr10 scaled \magstep 1
\font\ten=cmr10

\leftskip .25 in
\rightskip .25 in

\noindent{\bigggfnt  Newtonian approach  for the Kepler-Coulomb  } \Par
\noindent{\bigggfnt   problem from the point of view of velocity space } \Par
\noindent{\bigggfnt  } \Par

\vskip 6 pt

\noindent {\bigfnt \bf H.\ N.\ N\'u\~nez-Y\'epez}\footnote\dag{On sabbatical leave from Departamento de F\'{\i}sica, UAM-Izta\-palapa}\par

\noindent {\it Instituto de F\'{\i}sica `Luis Rivera Terrazas', Benem\'erita Universidad Aut\'onoma
 de Puebla, A\-par\-tado Postal J-48, C P 72570, Puebla  Pue., M\'exico }\Par
 \Par

\noindent {\bigfnt \bf E.\ Guillaum\'{\i}n-Espa\~na, A.\ Gonz\'alez-Villanueva, R.\ P.\ Mart\'{\i}nez y Ro\-mero\footnote\ddag{\rm On sabbatical leave from Facultad de Ciencias, UNAM}, A.\ L.\ Salas-Brito}\par
\noindent {\it Laboratorio de Sistemas Din\'amicos, Departamento de Ciencias B\'asicas, Universidad Aut\'onoma Metropolitana-Az\-ca\-pot\-zalco, Apar\-tado Postal 21-726,  C P 04000, Co\-yoa\-c\'an D.\ F., M\'e\-xico}\Par

\vskip 24pt

\centerline{\bigfnt Abstract}\Par

\noindent  The hodograph of the Kepler-Coulomb problem, that is, the path traced by its velocity vector, is shown to be  a circle and then it is used to investigate  the properties of the motion.  We obtain the  configuration space orbits of the problem starting from  initial
conditions given using nothing more than the methods of synthetic
geometry so close to Newton's approach. The method works with elliptic, parabolic and hyperbolic orbits; it can even be used to derive Rutherford's  relation from which the scattering cross section can be easily evaluated. We think our discussion is
both interesting and useful inasmuch as it serves  to relate the
initial conditions with the corresponding trajectories in a purely geometrical
way uncovering in the process some interesting connections seldom discussed in standard treatments.  \Par
\vfil
\eject
\centerline{\bigfnt Resumen}\Par

\noindent Demostramos que la hod\'ografa del problema de
Kepler-Coulomb, esto es, la trayectoria que sigue su vector velocidad, es una circunferencia y  la usamos para
establecer geom\'etricamente otras propiedades del movimiento. Obtenemos la relaci\'on entre la hod\'ografa  y la \'orbita en el espacio de las configuraciones para el problema de Kepler-Coulomb partiendo de condiciones iniciales dadas y em\-plean\-do nada m\'as que los m\'etodos de la geometr\'{\i}a sint\'etica tan caros a  Newton. El m\'etodo que proponemos incluye tanto a las \'orbitas el\'{\i}pticas, como a las parab\'olicas e hiperb\'olicas  y puede tambi\'en usarse para deducir la relaci\'on de Rutherford, la que es la  clave para obtener la secci\'on eficaz de dispersi\'on. Pensamos que nuestro enfoque es tanto interesante como \'util ya que  permite trazar las trayectorias usando s\'olo m\'etodos  geom\'etricos y, adem\'as, reconocer algunas relaciones que no son evidentes en los tratamientos m\'as usuales. 
\Par
\vskip 25 pt

\noindent Classification Numbers: 03.20.+i, 95.10.C\par

\vfill
\eject

\noindent{\bf 1. Introduction}\Par

\noindent  We have been analysing an approach to solve the Kepler-Coulomb problem employing  the properties of its hodograph and its relationship to their orbits [1--4]. Please recall that {\sl hodograph} is the name given to the path traced by the velocity vector of a  system in velocity space.  In this work we purport to express all our arguments in  geometric terms in a sort of Newtonian fashion. The hodograph as a mean towards understanding the dynamics of a system was introduced by Hamilton [5] ---he even invented the term--- during the last
century.  Hamilton was able to show that the hodograph under an inverse squared centre of force [5--7] or, as we call it here, in the
Kepler-Coulomb problem is always a circle. It is curious to notice that Hamilton proved  that, in a way, the ancient Greek astronomers were right, the motion of planets around the sun is indeed circular, they just got wrong the space since the hodograph inhabits velocity rather than configuration space. However, even if one knows that the
Kepler-Coulomb hodograph is circular in shape, it is natural to wonder how can
that circle be related  with the well-known orbits in
configuration space. Let us note that the problem is easily solved in an
analytical treatment since we can use the polar angular coordinate in the
plane of the orbit, $\theta$, for relating the trajectory in $v$-space
with the trajectory in $r$-space [2--4].  Furthermore, let us point out that the problem posed
has been already solved geometrically, since there exist  beautiful  methods 
developed by Maxwell [6] and by Feynman [8] to solve it.  \Par

  In this work we discuss a ---we expect--- novel geometric approach to the
relationship between the hodograph  and the orbit of the Kepler-Coulomb problem. We begin establishing the circular shape of the problem's hodograph using standard analytical methods and then rework the path from the hodograph to the orbit using techniques that ---we think--- are akin to those in the Principia [9]. In our view, this geometric approach uncovers  the geometrical beauty
associated with the physics of the problem which no doubt contributed
to the attraction felt towards it by many people trough the centuries, from the ancient Mayan astronomers to their modern counterparts. We do think this type of approach contributes to a better understanding of the interplay between the geometry and the physical properties of the solution to the Kepler-Coulomb problem. It is to be noted that, mostly, the
constructions presented here require no more than straight edge and
compass to be realized. But, before embarking in the discussion, let us convene that the trajectory in configuration space be always called the {\sl orbit} whereas the trajectory in velocity space
be  always called the {\sl hodograph}. \Par

\noindent{\bf 2. The hodograph of the Kepler-Coulomb problem}\Par

\noindent The equation of motion of a particle interacting with an inverse squared
centre of force is,

$$ m{d{\bf v}\over dt}=-{\alpha\over r^2} \er,   \eqn{1} $$

\noindent where $m$, $\alpha$, ${\bf v}$ and $\er$ are, respectively, the 
mass of the particle, a constant characterizing the interaction strenght 
(which is positive if the interaction is attractive or negative if it is 
repulsive), the velocity vector, and the unit vector in the  radial  
direction in configuration space. In the Kepler-Coulomb problem, described
by equation (1), the energy $E$ and the angular momentum ${\bf L}=m{\bf r}
\times {\bf v} =mr^2\dot \theta\ez= L\ez$ are conserved (we choose the
direction of ${\bf L}$ as $z$-axis).
The motion is thus seen to be confined to a plane orthogonal to
${\bf L}$.  In this orbital plane we may choose a polar coordinate
system with unit vectors $\er$ and $\eth=\ez \times \er$, for
describing the motion.  Given this information, showing that the
hodograph is a circle is not difficult, this has been done by Fano and 
Fano [10] using a nice but not completely geometrical approach. 
 As posing and solving dynamical problems in geometrical terms
is, however, very unfamiliar to modern readers, we have decided to 
start the discussion using conventional differential equation 
techniques; according to this, we begin by giving an standard proof of 
the hodograph's main properties [2]: \Par

\noindent If we  multiply (1) times $L$, the equation of motion becomes

$$ Lm {d {\bf v}\over dt} = - {\alpha \over r^2} (mr^2\dot \theta)
\er= \alpha m \dot\eth,  \eqn{2} $$

\noindent where we used $\dot\eth=-\dot\theta\er$. From (2), it must be
clear that the Hamilton vector [2,11]

$$ {\bf h} = { {\bf v} - {\alpha \over L} \eth } \eqn{3} $$

\noindent is a constant of motion in the Kepler-Coulomb problem. As can be
seen in this equation, the Hamilton vector is always parallel to the
velocity at pericentre ${\bf v}_p$ [4, 12]. The magnitude of  the Hamilton vector 

$$ h =  \sqrt{{2E \over m} + {\alpha^2 \over L^2}}, \eqn{4}$$

\noindent increases when the energy $E$ increases or when the magnitude
of the angular momentum $L$ decreases.  Moreover, as follows from (3),
the velocity lays in a circular arc  with radius $R_h\equiv
\alpha/L$ and whose centre is at the tip of ${\bf h}$ in velocity
space.  The Coulomb-Kepler {\sl hodograph} is  a
circle  whose centre  is  at the tip of ${\bf h}$ and, therefore,
the Hamilton vector defines a dynamical symmetry axis of the
hodograph ---dynamical symmetry in the sense that it is not only a
geometrical property, the interaction intervenes directly; for
comparison note that the rest of the diameters are just geometric symmetry axes. 
This property of the hodograph shows that the orbit has also a dynamical symmetry axis;
such axis is found by geometric means in  section 3 below. \Par

As the hodograph is a closed curve ---at least when it happens to be
the whole circle, \ie\ in precisely the case of bounded orbits--- then all the 
bounded orbits of the problem have to be
necessarily periodic. How are other features of the hodograph related
to the properties of the orbit? As we exhibit below in sections 4.1 to 4.5,
the geometric shape and the bounded or unbounded nature of the orbits
change according to where the $v$-space origin is positioned in relation
to the hodograph. Many of these features are discussed in modern language for
the case of an attractive interaction in [2] and for the scattering case in [3--4]. \Par

\noindent {\bf 3. From  the initial conditions to the hodograph}\Par

\noindent  If we are given the position ${\bf r}_0$ and the velocity
${\bf v}_0$ at a certain time $t_0$, how can we construct the Hamilton
vector and the hodograph? In this section we show how this can be done
using a very simple geometrical construction.  Before beginning with the geometrical construction, we first need to
calculate the lenght of the angular momentum vector: $L=m r_0 v_0 \sin
\delta$, where $0\leq\delta\leq\pi$ is the angle between the initial
position and velocity. But, $L$ is just the area
of the rectangle spanned by $r_0$ and the component of ${\bf v}_0$
orthogonal to ${\bf r}_0$ times $m$, that is, it is twice the areal
velocity of Kepler second law. We also need the `length' $R_h=\alpha/L$
---remember that this ratio has really dimensions of velocity. \Par

To understand the geometrical construction that follows it is convenient
to keep figure 1 in sight. Let the point $Q$ be the position of the
centre of force. Draw the line segment $\ol{QR}$ corresponding to the
initial position ${\bf r}_0$ (in fact, this is always the name given to the
line segment representing the initial position in all the discussions
that follow).  Extend the segment $\ol{QR}$ up to an arbitrary point $O$
---this just corresponds to choosing the origin in velocity space. From
the $v$-space origin $O$, draw the line segment $\ol{OP}$ corresponding
to ${\bf v}_0$ ($\ol{OP}$ is always the name of the line segment
representing the initial velocity in all the discussions that follow)
and draw, perpendicular to $\ol{QR}$, a line segment $\ol{OO'}$  of
lenght $R_h$ ---that is, we are drawing $-\alpha/L \, \eth$ (recall that
we defined $\eth=\ez\times \er$, where $\ez\equiv {\bf L}/L $). Notice
that the previous construction assumes both an attractive interaction and  
all the conventions mentioned.  To
analyse a repulsive interaction the point ${O'}$ had to be chosen in the
opposite direction (\ie\ in such case we should draw $+\alpha/L \, \eth$).
\Par

Using the parallelogram rule, sum  $\ol{OP}$ to  $\ol{OO'}$  to get
the point $C$. The line segment $\ol{OC}$ represents the Hamilton vector. 
Having obtained $\bf h$, to get the hodograph  draw, with centre at $C$, a 
circle  of radius $R_h$; this represents the hodograph. \Par

The above geometrical construction besides giving  ${\bf h}$ and the
hodograph tell us about the bounded or unbounded nature of the
orbit. It is only a matter of
noticing whether $O$ is located  inside the circle of the hodograph or
not; if it is  inside, the orbit is bounded and the energy has to be negative, 
if not, the orbit is unbounded and the energy
is positive. Figure 1 illustrates a case in which $O$ is inside,
that is, a motion with $E<0$.  What about the case $E=0$? As it is easy
to see from (4), or just from the continuity of the descriptions, this 
case only happens when $O$ sits precisely on the circle, that is, when
$h=R_h=\alpha/L$ [2]. \Par

 It is also easy to obtain the dynamical symmetry axis of the orbit
from the given initial conditions. We just need to draw the line
segment $\ol{QS}$, which is a line perpendicular to  $\ol{OC}$ passing
through the centre of force $Q$.  This follows from the paralellism of
${\bf h}$ and the velocity at pericentre ${\bf v}_p$. The line $\ol{QS}$ 
so drawn, {\sl is the orbit's dynamical symmetry axis}. Notice also that
${\bf v}_p$ can be drawn by simply prolonguing the segment $\ol{OC}$
until it intersects the hodograph. This intercept is marked $X$ in
figure 1. If, as happens in figure 1, there are two intersections with
the hodograph and not just one, the velocity space origin  $O$ is,
necessarily,  inside the hodograph, that is, the energy is necessarily
negative. The second intercept, labeled $X'$ in figure 1, defines the
segment $\ol{OX'}$ corresponding to the velocity at the {\sl apocentre}
of the orbit, that is, to the point on the orbit farthest from the
centre of force and therefore with the least magnitude. A such point
obviously does not exist in the $E>0$ case when $O$ is outside the
hodograph.\Par

 In all the section 4, we assume that the symmetry axis has been
drawn as described before the geometric discussion begins.  The
just found dynamical symmetry axis corresponds to the direction of the
Laplace-Runge-Lenz vector ${\gros{\cal A}}={\bf h}\times{\bf L}$ [13--14]
which always points towards the pericentre of the orbit. \Par

\noindent {\bf 4. From the hodograph to the orbit}\Par

\noindent In this section we  show how given the hodograph, constructed
from the initial conditions  as explained in section 3,   the orbit in
configuration space can be obtained and all its geometrical properties
established. \Par

\noindent {\sl 4.1 The case of an attractive interaction with the
$v$-origin inside the hodograph} \Par

\noindent Let us assume that the origin of coordinates in velocity space
is within the circle of the hodograph; this is the case whose realization from initial conditions was previously discussed in section 3 and was illustrated in figure 1.
Please refer to figure 2 for the schematic representation of the geometric 
steps that follow and as an aside note that every single step can be accomplished using  only
straight edge and compass. \Par

 The points $Q$, $R$, $O$, $O'$, $P$ and $C$ in
figure 2 have exactly the same meaning as in figure 1, that is, they
serve to construct  the Hamilton vector $\ol{OC}$ and the hodograph
centered at $C$ given the initial conditions ${\bf r}_0$ (the straight
line $\ol{QR}$) and ${\bf v}_0$ (the straight line $\ol{OP}$), and the
vector $-\eth R_h $ (the straight line $\ol{OO'}$). In fact,  we will 
always assume this meaning
for the labeling of points in figures 2--5, in figure 6 the naming of
points is similar excepting that $O'$ is not found to be necessary. \Par

To locate any point in the  orbit, extend the straight line
$\ol{PO}$ until it again intercepts the hodograph at point $T$
(see figure 2a).  Trace a perpendicular to $\ol{CT}$ passing through
the point $R$,  this line intercepts the symmetry axis (constructed as
in section 3) at the very important point $Q'$. To
locate the point on the orbit corresponding to any given point on the
hodograph, let us  notice that we have already one such pair of
points, the initial conditions: the points $R$ and $P$. Let us choose
another point $P'$ on the hodograph; to begin, draw the straight line
$\ol{OP'}$, extend it until it intersects the hodograph at point
$T'$. Draw  two straight lines perpendicular to $\ol{CP'}$ and to
$\ol{CT'}$ passing through $Q$ and $Q'$ respectively; we assert that
this two perpendiculars meet at the required point $R'$ on the orbit,
as was the case with the perpendiculars to the straight segments
$\ol{CP}$ and $\ol{CT}$, related to the initial conditions and meeting
at $R$. To draw the complete orbit, we have to repeat the same procedure
starting from each and every point on the hodograph, in this way
drawing, point by point, the whole orbit---which is shown as the gray curve which includes 
the points $R$ and $R'$ in figure 2a. \Par

What are the properties of the just constructed orbit? The easiest way
to answer this question is by establishing the orbital shape. To do this, let
us first draw the circular arc ${Q'W}$  centered at $R$ with a radius
equal to the lenght of $\ol{RQ'}$. This arc intercepts the straight line
$\ol{QO}$ at the point $W$ (see figure 2b).  Next, trace the circular arc
$WW'$ centered at $Q$ with radius $\ol{QW}$. It is now easy to see, just
by noticing that the  shaded triangles $\triangle P'T'C$ and
$\triangle W'Q'R'$ are both isosceles and similar to each other (this
happens by construction),  that the point $R'$ on the orbit is at the
same distance from the arc $WW'$ than from the point $Q'$. We can see
thus that the radius of the circular arc $WW'$ is the sum of the lenghts
of $\ol{QR'}$ and $\ol{Q'R'}$ and, therefore, that {\sl in the case
$E<0$ the orbit is necessarily an ellipse} with major axis $2a$ equal to
the lenght of the line segment $\ol{QW}$. The auxiliary point $Q'$ is
thus one of the foci of the ellipse, the other one coinciding with the
centre of force $Q$.  The line $\ol{QS}$  can be seen to be  the
symmetry axis of the ellipse as we have anticipated. In fact, the
eccentricity of the ellipse is easily calculated as
$\epsilon= h/R_h=OC/CP $ [2]. Thus,
the famous Laplace-Runge-Lenz vector can be drawn as a straight line
segment of lenght $\alpha\epsilon$ parallel to $\ol{SQ}$  ---as the
segment labeled $\cal{A}$ in figure 2a illustrates. It is to be noted that 
the circle $WW'W"$ can be identified with the circle
used by Maxwell and by Feynman in their respective discussions of the
Kepler-Coulomb problem [6,8].
\Par

\noindent {\sl 4.2 The case of an attractive interaction with the
$v$-origin on the hodograph} \Par

\noindent Let us now assume that the origin of coordinates in velocity
space happens to be precisely on the  circle of the hodograph, as shown
in figure 3. The symmetry axis $\ol{QS}$, as described in section
3, is the line perpendicular to $\ol{OC}$ which passes through the
point $Q$. For the construction, we also need the auxiliary line
$\ol{SW}$, paralell to $\ol{OC}$ and whose distance from the initial
point $R$ ($\ol{RW}$) is equal to the lenght of the segment $\ol{QR}$.
\Par

To construct the orbit, we first, just for the sake of convenience,
translate the centre of the hodograph to the point $Q$. That is, the
hodograph's centre is relocated to coincide with the centre of force.
See figure 3. All references to points on the hodograph from now on, assume this new location for it.  Let us choose an arbitrary point $P'$ on the hodograph
and draw the straight line segments $\ol{OP'}$ (the velocity) and
$\ol{QP'}$ (\ie\ the vector $\eth R_h$). Draw a perpendicular to
$\ol{OP'}$ passing through $Q$ and intercepting the auxiliary line
$\ol{SW}$ at the point $W'$. Erect $\ol{W'R'}$ perpendicular to
$\ol{SW}$ and trace $\ol{QR'}$ perpendicular to $\ol{QP'}$ passing
through $Q$. This line  intercepts $\ol{W'R'}$ at $R'$, a point on the
orbit. We assert that any point constructed in this way belongs to a
parabola which thus corresponds to the shape of the orbit in the case in
which the $v$-origin sits on the hodograph, that is, in the $E=0$ case.
In fact, the assertion can be checked just by noting that the initial
conditions are related in exactly the same way as we did in the
previous section. \Par

The proof that the orbit is a parabola is similar to that given for the
elliptic case of subsection 4.1, based as it is on the
similarity of the shaded isosceles triangles $\triangle QP'O$ and
$\triangle R'QW'$ in figure 3 (please remember that we are always referring to points in the displaced (continuous) hodograph). This similarity is enough to show that
the lenghts of $\ol{QR'}$ and of $\ol{R'W'}$ are the same, thus
establising the orbit as the locus of points equidistant from both the
point $Q$ and the straight line $\ol{SW}$. Therefore, $Q$ is seen to be
the focus and the segment $\ol{SW}$ the directrix of the parabola.
Notice that the directrix is defined by the direction of the Hamilton
vector ${\bf h}$, being thus also parallel to the velocity at pericentre
$\ol{CX}$ (${\bf v}_p$). Notice also that both the hodograph and the
orbit exhibit that the speed at pericentre (the lenght of $\ol{CX}$,
$v_p$) is always greater  than any other speed in the problem. \Par

\noindent {\sl 4.3 The case of an attractive interaction with the
$v$-origin outside the hodograph} \Par

\noindent Let us now assume that the origin of coordinates in velocity
space $O$ is outside the circle of the hodograph as shown in figure 4.
The Hamilton vector is $\ol{OC}$ and the hodograph is the circle
centered at $C$ with radius $\ol{CP}$.   A very important difference
with the cases of subsections 4.1 and 4.2, is that here the hodograph
is not the whole circle since, in this case, this is the only way of
guaranteeing that every point in the orbit came from just one and only
one point, \ie\ every point corresponds to just one velocity. In this
way we also guarantee that every other speed $v$ is always less than the
speed at pericentre $v_p$ [3--4].
This implies that, in the attractive case considered here, the hodograph
is  the arc of the circle `farthest' from the origin ---shown as a
continuous line in figure 4a. As in the previous two subsections, the
symmetry axis $\ol{QS}$, is the line perpendicular to $\ol{OC}$ and
passing through the centre of force  $Q$ as illustrated in figure 4a.
\Par

For constructing the orbit, we first need to locate the auxiliary point
$Q'$ (figure 4a). To locate $Q'$ first trace the
straight line $\ol{CT}$, were $T$ is the unphysical intercept of the
hodograph with the line $\ol{OP}$, then erect on $R$ a  perpendicular to
$\ol{CT}$. The intercept of this last line with the symmetry axis
$\ol{QS}$ is precisely the auxiliary point $Q'$. With these geometric
data we can begin the construction of the orbit. \Par

Let us select any point $P'$ on the hodograph, trace the straight lines
$\ol{CP'}$ and $\ol{CT'}$.
Erect perpendiculars to them passing through $Q$ and $Q'$, respectively,
the intersection of these lines is another point $R'$ on the orbit. It
is now obvious that for constructing the whole orbit you have to repeat
this procedure over and over again, starting from each point on the
hodograph, you can check that the initial conditions are related by this
same procedure.  The asymptotic velocities and the speed at infinity are
also easy to obtain. To this end just trace, starting  from the
$v$-origin $O$, the straight line segments, $\ol{OB}$ and $\ol{OB'}$,
tangent to the hodograph. These segments correspond, respectively, to
the asymptotic velocities ${\bf v}_{-\infty}$ and ${\bf v}_{\infty}$
(as follows from angular momentum conservation), therefore,  as can be
seen in figure 4b, their common lenght is the sought after speed
$v_\infty=\sqrt{h^2-R_h^2}$. \Par

We are just left with the task of establishing the shape of the orbit.
To this end, trace the circular arc $QW$ centered at $R$ with radius
$QR$, this arc intercepts the segment $\ol{Q'R}$ at the point $W$. See
figure 4b. Next, trace a circle centered at the auxiliary point $Q'$
and radius $Q'W$. Given  the similarity of the two shaded isosceles
triangles $\triangle CP'T'$ and $\triangle R'QW'$, we can assert
that any point $R'$  is at the same distance from $Q$ and from the
circle $WW'$ (shown as a continuous dark circle in figure 4b), where
$W'$ is the intercept of this last circle with $\ol{Q'R'}$. From the
fact that any point on the orbit is at the same distance from the point
$Q$ and from the circle $WW'$, we can establish that the difference
between the distances from $R'$ to $Q$ and from $R'$ to $Q'$, is
equal to the radius of the circle $WW'$ and, therefore, it is a
constant. But this is precisely the definition of an hyperbola, which
is thus the shape of the orbit in the $E>0$ attractive case. This is
illustrated in figure 4b.
\Par

\noindent {\sl 4.4 The case of a repulsive interaction} \Par

\noindent In the previous sections we have been addressing the
construction of orbits in the case of an attractive interaction in
equation (1), \ie\ the case with $\alpha>0$; however, the sign of
$\alpha$ does not really matter for the shape of the hodograph, it is
{\sl always} a circle. But, as we already know [13], there are nevertheless  
differences in the kind of motions
in configuration space that are allowed. How can we understand such
differences starting from just the hodograph? Finding a sort of
geometric answer to this question is one of the purposes of this section.
\Par

Notice that, in the case at hand and as shown in figure 5a, both points
$O$ and $P$ are on the same side of the straight line segment $\ol{QR}$,
therefore the length of $\ol{OL}$ (the Hamilton vector ${\bf h}$) is
 greater than $\ol{OP}$ (the initial velocity ${\bf v}_0$) and that
$\ol{OO'}$ (the hodograph radius is less than $\ol{OP}$, this means
that the origin of coordinates in velocity space is always outside the
 circle of the hodograph. That is, whenever $\alpha <0$ and since ${\bf v}_0\cdot \eth$
 could not be negative nor vanish, the only possibility for the $v$-space origin
 is to be outside the hodograph. In the `modern' language of classical mechanics,
 if $\alpha <0$ then the only possible motions have a necessarily positive energy. \Par

 The points $Q$, $R$, $O$, $O'$, $P$, $T$ and $C$ in figure 5 have exactly the same meaning
 as in the previous figures 1 to 4, that is, they serve to construct the Hamilton vector $\ol
 {OC}$ and the hodograph centered at $C$, given the initial conditions {\bf r}$_0$, the straight
 line $\ol {QR}$, and {\bf v}$_0$, the straight line $\ol {OP}$, and the vector $- \hat
 {\bf e}_\theta R_h$, represented by the straight line $\ol {OO'}$, This case is similar to that
 of section 4.3 since the hodograph is not the whole circle; this can be argued using essentially
 the same argument as in that section [2--4]. In the
 repulsive case considered here the hodograph is the circular arc     
`closer' to the origin ---which is shown as a continuous line in
figure 5. As in the previous subsections, the symmetry axis
$\ol{QS}$ is the line perpendicular to $\ol{OC}$ and passing through
the centre of force  $Q$.  \Par

To find the orbital shape we need the auxiliary point $Q'$, which is the
intercept of a perpendicular to $\ol{CT}$ going through $R$ with the
symmetry axis $\ol{QS}$. Now is just a matter of choosing an arbitrary
point $P'$ on the hodograph, and prolonguing the straight line segment
$\ol{OP'}$  until it again meet the hodograph  at point $T'$. Trace the
straight line segments $\ol{CP'}$ and $\ol{CT'}$ and erect perpendicular
segments going through $Q$ and $Q'$, respectively.  The intercept of these
perpendiculars is the corresponding point $R'$ on the orbit. Repeating the
procedure for every point on the hodograph we can obtain the whole orbit.
The orbit is again, as in section 4.4, an hyperbola with foci $Q$ and $Q'$, as can be shown by
considering that any point on the orbit is at the same distance from the
fixed point $Q'$ and from the auxiliary circle $WW'$ centered at $Q$, defined as in section 4.4.  The complete argument uses the similar isosceles triangles $\triangle CP'T'$ and $\triangle R'W'Q'$
and essentially repeats the argument of the previous section. \Par

\noindent {\sl 4.5  The Rutherford problem} \Par

\noindent Let us pick the point $Q$ as the location of the repulsive centre
of force. To describe geometrically a scattering situation, we have
basically the same situation of sections 4.3 and 4.4, the only difference
being that, here we are given the velocity ${\bf v}_{-\infty}$, \ie\ the
velocity evaluated at a time in `the infinitely distant past' and the
impact parameter $b$,  not  the velocity and the position at a certain
finite time $t$.  See figure 6. With the data just mentioned and from the  
location of the
centre of force $Q$, draw the line segment $\ol{OK}$ parallel to
${\bf v}_{-\infty}$, starting from the arbitrary point $O$ but passing at
a distance $b$ off the centre of force.  \Par

If on $\ol{OK}$ we choose the segment $\ol{OB}$ to represent 
${\bf v}_{-\infty}$, the point $O$ would have  been implicitly selected 
to play the role of the $v$-origin. Then, from the point $B$, erect a 
perpendicular straight line segment, of lenght
$R_h$, up to the point $C$. Next, centered at $C$ draw a circle with radius
$\ol{CB}$, a part of this circle is the hodograph of the problem. If we
draw the tangent to the circle $\ol{OB'}$, this represents the asymptotic
outgoing velocity at infinity ${\bf v}_{\infty}$; the hodograph is thus the
circular arc $BPB'$ and the Hamilton vector is the line segment $\ol{OC}$
bisecting the angle $\angle B'OB$. This
angle is usually called the deflection angle $\xi$. In fact, the right
triangle $\triangle OBC$ gives immediately the Rutherford relation between
$\xi/2$ and $L$

$$ \tan {\xi\over 2}={R_h \over \sqrt{h^2-R_h^2}}={\alpha\over
v_{{}_{-\infty} L}},  \eqn {6} $$

\noindent which can be used as the starting point to derive the famous
Rutherford scattering formula [3,8,13].  See also [4] where the 
Rutherford problem is discussed taking a velocity space point of view from the start.   \Par

\noindent {\bf 5. Conclusions} \Par

\noindent We have exhibited that the orbits of the Kepler-Coulomb problem can
be obtained and classified (basically in terms of the energy) starting from
the hodograph and using techniques of synthetic geometry  requiring no more
than straight edge and compass. We have exhibitted that  the Hamilton vector is
crucial for deciding geometrically if the orbits are bounded or not and,
furthermore, that with its help, we can draw point by point any  orbit whatsoever. On the other 
hand, speaking on the purely geometrical content of the
paper, we have managed to offer an admittedly not very systematic proof
of an elementary but not widely known geometric  result, namely, that
the conic sections can be defined as the locus of points equidistant from
both a fixed point and a fixed circle.  The geometric  method can be further justified as in [15].\Par

We have learnt a lot in trying to do mechanics  using the nowadays non-standard methods of Newton. We hope that this article may convey to the readers the aesthetic pleasures we discovered in the geometric structure of  Newton's mechanics. We think these considerations are enough to justify the  approach presented in this article which  exhibit the enormous power of geometric reasoning in classical mechanics [16].  However, we have to emphasize that Newton's geometric methods go  far beyond the simple results obtained here; it has been discovered, for example, that the Principia contains, among other things, astonishing geometric proofs of deep results on the properties of Abelian integrals [17]. \Par

\noindent{\bf Acknowledgements}\Par

\noindent  This work has been partially supported by CONACyT (grant 1343P-E9607). ALSB and HNNY 
want to  acknowledge  very interesting discussions with Leticia Fuchs G\'omez. We must also thank  our colleagues D.\ Moreno, J.\ A.\ Gonz\'alez Men\'endez, and  R.\ W.\ G\'omez Gonz\'alez for their very useful comments. This paper is dedicated with thanks to Q.\ Kuro, Q.\ Sieri, M.\ Tlahui, M.\ Miztli, G.\ Tigga,  B.\ Caro, Ch.\ Ujaya, U.\ Sasi and F.\ C.\ Minina. AGV wants to express his  thanks to Luis Gonz\'alez y Gonz\'alez and Armida de la Vara for all the support and encouragement they have always given him. We should also thank a careful referee that pointed out  several misprints and omissions  in the original version of the article and that shared with us our love for geometric argumewnts.\Par

\vskip 15 pt
\vfil
\eject

\noindent{\bf References} \Par
\vskip 4 pt

{\ten \baselineskip 11.5 pt

 \noindent [1] O.\ Campuzano-Cardona, H.\ N.\ N\'u\~nez-Y\'epez, A.\ L.\ Salas-Brito and
 G.\ I.\ S\'anchez-Ortiz,  {\it Eur.\ J.\ Phys.\ } {\bf 16} (1995)  220. \Par

\noindent  [2] A.\ Gonz\'alez-Villanueva, H.\ N.\  N\'u\~nez-Y\'epez and
A.\ L.\ Salas-Brito,   {\it Eur.\ J.\ Phys.\ } {\bf 17} (1996) 168.  \Par

\noindent [3] A.\ Gonz\'alez-Villanueva, H.\ N.\ N\'u\~nez-Y\'epez and  A.\ L.\ Salas-Brito,  {\it Rev.\ Mex.\ Fis.\ } {\bf 44} (1998) 183.\Par

 \noindent  [4] A.\ Gonz\'alez-Villanueva, E.\ Guillaum\'{\i}n-Espa\~na,
H.\ N.\ N\'u\~nez-Y\'epez and  A.\ L.\ Salas-Brito,   {\it Rev.\ Mex.\ Fis.\ } {\bf 44} (1998)
 303.\Par

 \noindent [5] W.\ R.\ Hamilton,  {\it Proc.\ Roy.\ Irish Acad.\ } {\bf 3} (1846)
344.\Par

\noindent [6] J.\ C.\ Maxwell {\it Matter and motion}, New York,
Dover (1952) 107. \Par

\noindent [7] W.\ Thomson  and P.\ G.\ Tait, {\it Treatise on natural philosophy},
  New York, Dover (1962) \S 37--\S38. \Par

\noindent [8] D.\ L.\ Goodstein and  J.\ R.\ Goodstein,  {\it Feynman's lost
lecture. The motion of planets around the sun}, New York, Norton (1996) Ch 4.\Par

\noindent [9] I.\ Newton,  {\it Principios matem\'aticos de la filosof\1ia
natural},  Bar\-celona, Altaya (1993) Libro I. \Par

\noindent [10] U.\ Fano  and L.\ Fano,  {\it Basic physics of atoms and molecules}, New York,
John Wiley  (1959) Appendix III. \Par

\noindent  [11] D. Moreno,   {\it Gravitaci\'on Newtoniana}, M\'exico D.\ F., FCUNAM (1990). \Par

 \noindent [12] R.\ P.\ Mart\'{\i}nez-y-Romero, H.\ N.\ N\'u\~nez-Y\'epez, and
A.\ L.\ Salas-Brito,  {\it Eur.\ J.\ Phys.\ } {\bf 14} (1993) 71. \Par

 \noindent [13] L.\ Landau and E.\ M.\ Lifshitz,  {\it Mechanics}, Pergamon, Oxford (1976), Chap.\ IV.\Par

\noindent [14] A.\ L.\ Salas-Brito, R.\ P.\ Mart\'{\i}nez-y-Romero, H.\ N.\ N\'u\~nez-Y\'epez,  {\it Intl.\ J.\ Mod.\ Phys.\ A} {\bf 12} (1997) 271.  \Par

\noindent [15] A.\ Gonz\'alez-Villanueva, E.\ Guillaum\1in-Espa\~na, R.\ P.\ Mart\1inez-y-Romero, H.\ N.\ N\'u\~nez-Y\'epez, A.\ L.\ Salas-Brito, {\it Eur.\ J.\ Phys.\ } (1998) in press. \Par

\noindent [16] T.\ Frankel,  {\it The geometry of physics,} Cambridge University Press, Cambridge (1997).\Par

\noindent [17] T.\ Needham, {\sl Am.\ Math.\ Month.} {\bf 100} (1993) 119. \Par

 \vfill
 \eject

\baselineskip 13 pt

\centerline{\biggfnt Figure Captions}\Par

\noindent Figure 1\par

\noindent The geometrical procedure for obtaining the Hamilton vector
and the hodograph from given initial conditions ${\bf r}_0$ and ${\bf v}_0$
is illustrated. $O$ labels the origin of coordinates in velocity space or
$v$-origin and $Q$ labels the location of the centre of force. To draw the
segment ${OO'}$, corresponding to $-\eth\alpha/L$, we assumed that
${\bf L}$ points outside the plane of the paper. The Hamilton vector is the
line segment $\ol{OC}$, the circle  $X'PX$ centered at $C$ is the
hodograph. The straight line segments $\ol{SX}$ and $\ol{QS}$ correspond,
respectively,  to the dynamical symmetry axis of the orbit and of the
hodograph. The discussion related to this figure can be found in section
3.\Par
\vskip 5 pt

\noindent Figure 2\par

 \noindent 2a.\ The procedure for reconstructing the orbit when the hodograph
encompass the $v$-origin is illustrated. The meaning of the points marked $Q$, $R$, $O$, $O'$, $P$ and $C$ is illustrated. Notice that the orbit is indeed closed; furthermore, notice that despite appearances the point $S$ does {\bf not} necessarily correspond to the vertex of the ellipse.  For the detailed discussion of this case  see the section
4.1. \Par
\noindent 2b.\ To demonstrate the orbit is indeed an ellipse we need to
recognize that the two shaded isosceles triangles $\triangle Q'R'W'$ and $
\triangle CP'T'$ are similar to each other. \Par

\vskip 5 pt

\noindent Figure 3\par

\noindent  The procedure for reconstructing the orbit when the $v$-origin
is precisely on the hodograph is illustrated. For the sake of convenience, let us first 
translate the whole hodograph from its original place centered at $C$ to a new location centered at  $Q$ (the centre of force). {\sl All references to points on the hodograph are to be understood at its  displaced location}.
To demonstrate that the  orbit
is a parabola we only need to recognize that the two shaded isosceles
triangles $\triangle QP'O$ and $\triangle R'QW'$ are similar to each other.
Notice that the straight line segment $SW'W$ corresponds to the auxiliary
circle of the previous figure. Thus, from this point of view, the directrix
is just a degenerate circle with infinite radius. See the discussion in
section 4.2.\Par

\vskip 5 pt
\noindent Figure 4\par

\noindent 4a.\ The procedure for reconstructing the orbit when the $v$-origin
is outside the hodograph  is illustrated. The points $P$ and $P'$ on the hodograph 
correspond to the points $R$ and $R'$ on the orbit.See section 4.3 \Par

\noindent 4b.\ To demonstrate that the orbit is an hyperbola whose internal
focus coincides with the centre of force, we only need to recognize that the
two shaded isosceles triangles $\triangle CP'T'$ and $\triangle R'QW'$ are
similar to each other. \Par

\noindent

\vskip 5 pt

\noindent Figure 5\par

 \noindent  The procedure for reconstructing the orbit starting with finite
initial conditions. Here we consider the case when the $v$-origin is
outside the hodograph  and the interaction is repulsive.
 The two shaded triangles are important for the discussion in section 4.4.
\Par

\vskip 5 pt

\noindent Figure 6\par

\ni  The Rutherford relation between $\xi$ and $L$ can be simply obtained from the hodograph as we illustrate in this figure. We also exhibit the procedure for reconstructing the orbit in a {\sl scattering} situation. We consider the case when both the $v$-origin is outside the
hodograph (\ie\ the case $E>0$) and the interaction is repulsive (\ie\ $\alpha <0$). 
For a brief discussion see section 4.5. A complete discussion from the point of view of velocity
space can be found in Ref.\ 4.
 \Par
}
 \vfil
 \eject
 
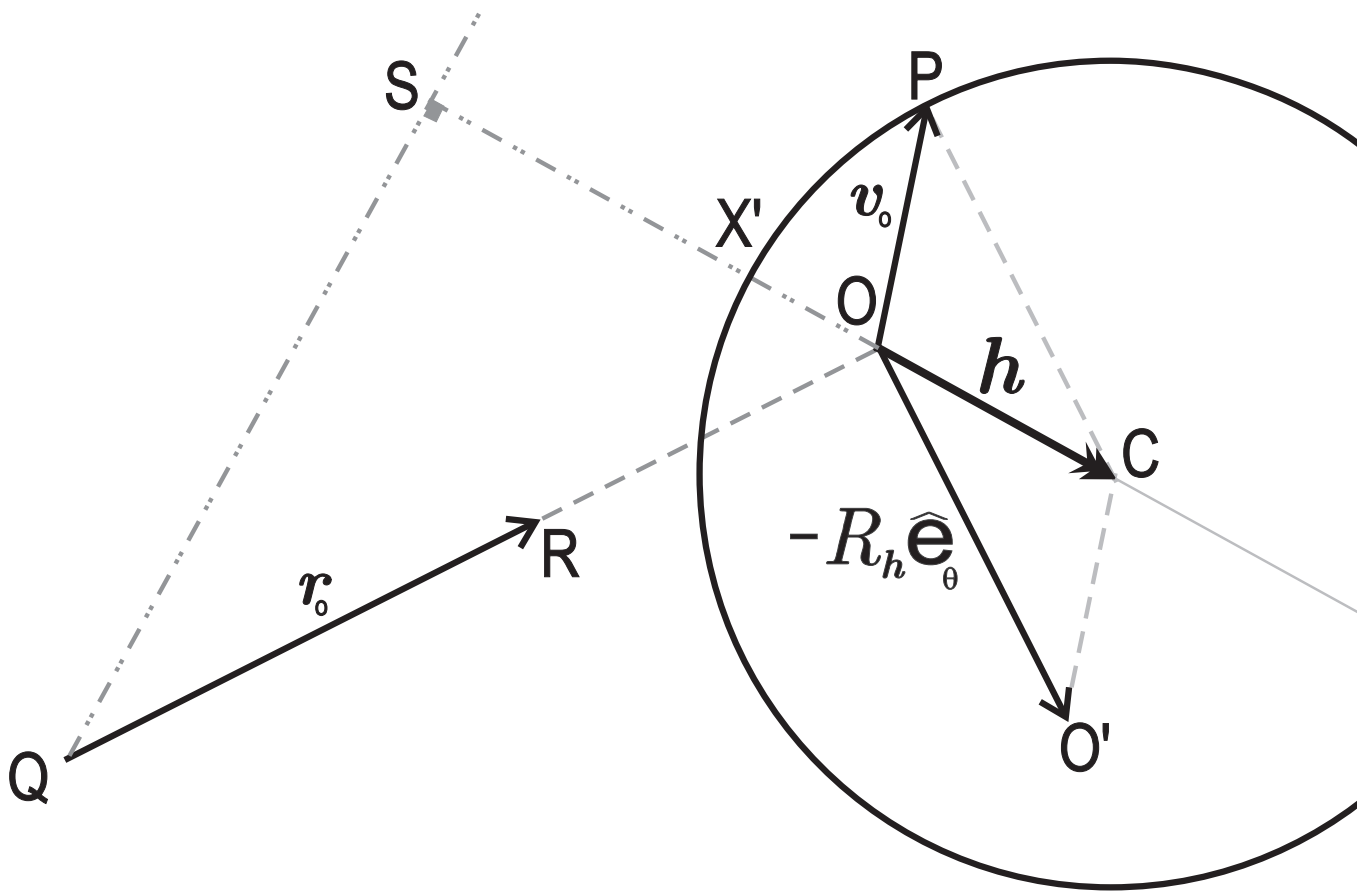
\end